\documentclass{amsproc}
\usepackage{amssymb}

\usepackage{graphicx}
\usepackage{amscd}

\theoremstyle{plain}

\numberwithin{equation}{section}

\input{tcilatex}

\begin{document}
\title{Quasi-stationary Stefan problem and computer simulation of interface dynamics.}
\author{R.\ Andrushkiw}
\address{New Jersey Institute of Technology University Heights Newark, NJ 07102-1982 }
\email{roandr@megahertz.njit.edu }
\urladdr{http://eies.njit.edu/\symbol{126}andrushk/}
\author{V. Gafiychuk}
\address{Institute of Applied Problems of Mechanics and Mathematics, National Academy
of Sciences of Ukraine, Lviv}
\email{viva@iapmm.lviv.ua}
\author{R. Zabrodsky}
\address{Institute of Applied Problems of Mechanics and Mathematics, National Academy
of Sciences of Ukraine, Lviv}
\email{viva@iapmm.lviv.ua}
\date{November 26, \ 2001}

\begin{abstract}
The computer simulation of quasistationary Stefan problem has been realized.
Different representations of the Laplacian growth model are considered. The
main attention has been paid for the interface dynamics represented by
integro differential equations. Numerical approach has been realized by use
of interpolating polynomials and exact quadrature formulae. As a result
system of ordinary differential equations has been simulated.
\end{abstract}

\maketitle

There are various physical processes in nature, such as crystallization,
combustion, deposition etc., which make it is possible to single out moving
boundary having physical properties different on its either side. The
location of the given boundary is not known beforehand and is determined
self-consistently by the distribution of its physical parameters in the
ambient space. Such problems as usually are singled out as a separate
section of modern mathematical physics, where the problems with the moving
boundaries, also known as Stefan problems, are considered. The number of
publications, devoted to this problem, is rather great. Therefore, we will
mention only some monographs \cite{mb2001,fb1999} including a lot of
bibliographies on mathematical models of various physical systems containing
moving boundaries.

Such problems, as usual, have no analytical solution due to the essential
non-linearity stipulated by representation of boundary conditions for
selfconsistent parameters allocated on the free boundary. Though it is
sometimes\ possible to find an analytically self-similar solution to a
Stefan problem in one-dimensional or quasi-one-dimensional case, concerning
the two-dimensional case such problems can be solved using computer
simulation only. Due to the fact that problems with the moving boundaries
describe a wide class of important technological processes, such as
crystallization, melting, etching, oxidation, deposition, diffusion etc.,
development of numerical methods of their solution is especially important 
\cite{mb2001,fb1999}.

\section{Model}

Let us consider crystallization of an undercooled melt located in infinite
container, with initial temperature $T<T_{c}$, where $T_{c}$ is the
temperature of crystallization \cite{65}. It is supposed that growth of the
interface of the solid and liquid phases $\Gamma $ is controlled by the
extract of heat generated in the process of crystallization inside the melt
(Fig. \ref{f11n}). Then the velocity $v_{n}$ of the motion of the
crystallization front along the normal vector $\mathbf{n}\;$to the\
interface $\Gamma $ is given by the expression:

\begin{equation}
v_{n}=-\frac{\varkappa }{\rho L}(\bigtriangledown T|_{\Gamma }\mathbf{n}).
\label{B.1}
\end{equation}
In the melt the distribution of temperature $T$ is determined by the heat
equation:

\begin{equation}
c_{\rho }\rho \frac{\partial T}{\partial t}=\mathbf{\varkappa \triangle }T.
\label{B.2}
\end{equation}
Here $c_{\rho }$ and $\rho $ are the specific heat and density of the liquid
phase; $\varkappa $ is thermal conductivity; $L$ is the latent heat of
crystallization. The temperature $T_{s}$ at the crystallization front
usually satisfies the Gibbs-Thomson\ condition:

\begin{equation}
T_{s}=T_{c}(1-Kd_{0}),  \label{B.3}
\end{equation}
where $K$ is the local curvature of the surface; $d_{0}$ is the capillary
length proportional to the surface energy.

\FRAME{ftbpFU}{1.6034in}{1.6898in}{0pt}{\Qcb{A schematic view of the
interface during crystallization.}}{\Qlb{f11n}}{f11n.gif}{\special{language
"Scientific Word";type "GRAPHIC";maintain-aspect-ratio TRUE;display
"USEDEF";valid_file "F";width 1.6034in;height 1.6898in;depth
0pt;original-width 2.4794in;original-height 2.6143in;cropleft "0";croptop
"1";cropright "1";cropbottom "0";filename 'F11n.gif';file-properties
"XNPEU";}}

It is known that in systems described by equations (\ref{B.1}) -- (\ref{B.3}%
) regular dynamics of the interface is unstable. The basic issues for
understanding the processes of dissipative structures formation during
crystallization are presented in \cite{MS64,MS63}, clarifying the reason of
formation of these structures for instance the morphological instability. In
case of the undercooled melt the mechanism of this instability is stipulated
by magnification of the motion velocity of convex segments of the surface
penetrating into the undercooled region. The latter stimulates their faster
propagating into depth of the melt.

The realization of instability of the homogeneous state of an interface for
crystallization of the undercooled melt, within a wide interval of wave
numbers and the essential instability of the whole process of shaping,
reduces itself to large varieties of possible structures and next to the
reproduction of the self-similar forms of the interface in smaller scales.
It also indicates the fractal character of their formation. In spite of
apparent simplicity of the problem statement, even in two-dimensional
domains it is often senseless to perform direct numerical calculations for
solving the problem.

In same cases the microscopic temporal scales of the parameters specifying
the problem changes can appear to be considerably smaller than the temporal
scales of dissipation and distribution of an appropriate unlocal field. As a
result the distribution of the unlocal field can be considered
quasistationary subject to the location of the boundary. In this case one
can construct the equations of surface motion containing only undefined
fields specified on the considered surface. It is the basic methodology for
this approach.

The essence of the method devised in \cite{65,68} implies replacement of a
consistent system of equations of motion by means of more simple ones, while
keeping all the properties of the origin problem. As the velocity in the
obtained equations is defined locally at each point, we obtain a set of
equations of local dynamics. Some simplifications, however, restricted
application of the approach proposed. We can see that the equations obtained
in \cite{65} are satisfied for undercooling, when the characteristic scale
of an unlocal field (for example temperature) $l_{t}=\varkappa \left( c\rho
v\right) ^{-1}$ is much smaller than the radius of surface curvature, $r$
i.e.$:l_{t}\ll r$. Although in this case one can not obtain solutions
without numerical simulation making use of appropriate approximations allows
to find new solutions, for example those in the form similar to snowflake.

If $l_{t}\gg r$ , then in the field of spatial changes of the temperature it
is possible to allocate two characteristic subregions. The first one, with
characteristic scale, comparable to $l_{t}$, corresponds to the change of
the temperature, due to the effect of thermal conduction, when the curvature
of the boundary element can be neglected. The second subregion lies in
proximity to the boundary element, and its thickness is equal to the average
curvature radius $r$. The change of the temperature and its gradient in this
subregion depends on curvature of the boundary. The modulation of the
surface temperature and its spatial distribution, as the first approximation
in the parameter $r/l_{t}$, corresponds to the Laplace equation. Since
harmonic functions as the solution of Laplace equation describing the
distribution of temperature, methods of complex analysis appear to be very
effective. This book presents methods of conformal mappings for analysis of
motion of boundary surface at least in two-dimensional case. Thus, if $%
l_{t}\gg r$ the problem of crystallization of undercooled melt (\ref{B.1})
-- (\ref{B.3}) and the problems considered above are formally equivalent and
are typical representatives of the quasistationary Stefan problem. Making
use of methods of complex variables allows us to reduce the problem
discussed to a system of ordinary differential equations \cite{ben,ben1,71}.

In this case problem (\ref{B.1}) -- (\ref{B.3}) becomes one of the simplest
among the ones considered above and is a kind of aquasistationary Stefan
problem. However, even in this simple case, determination of the boundary is
a rather complicated mathematical problem.

Reformulate now the problem (\ref{B.1}) -- (\ref{B.3}) in terms of Laplacian
growth problem. We investigate the growth of the two-dimensional region $Q$
and suppose that its boundary is $\Gamma $, which represents the physical
interface (Fig. \ref{f11n}). The field $\varphi $ outside the region $Q$,
satisfies the Laplace equation 
\begin{equation}
\Delta \varphi =0.  \label{f1}
\end{equation}
The boundary $\Gamma $ grows at a rate, that is proportional to the normal
gradient of the field of the interface. Therefore the evolution of the
interface is governed by the following equation 
\begin{equation}
\upsilon _{n}=\left. -\nabla _{\mathbf{n}}\varphi \right| _{\Gamma }.
\label{f2}
\end{equation}
The potential field $\varphi $ satisfies the following boundary condition on
the interface 
\begin{equation}
\left. \varphi \right| _{\Gamma }=\left. d_{0}k\right| _{\Gamma },
\label{f3}
\end{equation}
where $k$ is the curvature of the interface $\Gamma $, and $d_{0}$ is the
dimensionless surface tension parameter.

For this type of system the utilization of the theory of function of complex
variable is very fruitful. Theoretical methods of the functions of complex
variable appear to be rather useful in the construction of the effective
surface dynamics of the interface at least in the two-dimensional problem.
In this case the central issue of this problem belongs to the conformal
mappings of the physical regions of the space onto some special kind of
regions. A large number of models for interfacial dynamics had been proposed
up to now. Reviews of these problems can be found in \cite{pec,ben1,glb}.
The most famous equation was introduced by Shraiman and Bensimon \cite{sb}.
It is widely known as the conformal mapping equation and is still under
investigation.

In examples of the quasistationary Stefan problem, the main point of its
solution lies in construction of an appropriate potential specified by its
boundary value. Physically it is possible to do only when the boundary of
the domain of the potential is rather simple. The simplest examples of such
region is a disk. If we know a conformal mapping of a disk on some region,
the solution to the posed boundary problem reduces to change of variables in
the solution for the disk. Thus, solution of the quasistationary Stefan
problem for the two-dimensional region is practically reduced to
determination of an appropriate conformal mapping or its equation of motion.
It is possible to compare the non-stationary Stefan problem with an
equivalent quasistationary Stefan problem. It allows us to solve a wider
class of problems in framework of the equations of motion of the appropriate
conformal mappings.

Let us consider the approach developed by \ Shraiman and Bensimon \cite
{ben1,glb,sb,ben,gl1}. This approach is based on the Riemann mapping
principle. This principle states the existence of a conformal map from the
exterior of domain $Q$ onto the standard domain, for example, the exterior
of unit disk, with the boundary $\Gamma $ corresponding to unit circle. In
the outline under regard one can parametrize evolution of the interface in $%
z $ plane by means of time-dependent conformal map $F\left( w,t\right) $
which takes the exterior of the unit disk at each moment of time, $\left|
w\right| \geq 1$, onto the exterior of $Q$. Therefore the evolution of the
interface can be presented as follows: 
\begin{equation}
\Gamma \left( \theta ,t\right) =\lim_{w\rightarrow e^{i\theta }}F\left(
w,t\right) ,\text{ }0\leq \theta <2\pi .  \label{f4}
\end{equation}
As it is shown by Shraiman and Bensimon \cite{sb}, in the case of free
surface tension $\left( d_{0}=0\right) $ the boundary map satisfies the
following equation of motion 
\begin{equation}
\frac{\partial \Gamma \left( \theta ,t\right) }{\partial t}=-i\frac{\partial
\Gamma \left( \theta ,t\right) }{\partial \theta }\left. S\left[ \left| 
\frac{\partial \Gamma \left( \theta ,t\right) }{\partial \theta }\right|
^{-2}\right] \right| _{w=e^{i\theta }},  \label{f5}
\end{equation}
where $S\left[ \cdot \right] $ means the Schwartz operator \cite{shwarz}.
For the exterior of a unit circle with some boundary condition $f\left(
\theta \right) $\ the Schwartz operator can be presented as follows: 
\begin{equation}
S\left[ f\left( \theta \right) \right] =-\frac{1}{2\pi }\int\limits_{0}^{2%
\pi }d\theta ^{\prime }\frac{e^{i\theta ^{\prime }}+w}{e^{i\theta ^{\prime
}}-w}f\left( \theta ^{\prime }\right) +iC,  \label{sw}
\end{equation}
where $C$ is an arbitrary constant.

It was shown\cite{gsd} that equation (\ref{f5}) is equivalent to equation of
motion of certain field $\eta \left( \theta ,t\right) =\left| \frac{\partial
\Gamma \left( \theta ,t\right) }{\partial \theta }\right| :$

\begin{gather}
\frac{\partial \eta (\theta ,t)}{\partial t}=\frac{1}{R(t)}\frac{\partial
\eta (\theta ,t)}{\partial \theta }\frac{1}{2\pi }\int\limits_{0}^{2\pi
}d\theta ^{\prime }v_{n}(\theta ^{\prime },t)\eta (\theta ^{\prime },t)\cot 
\frac{\theta ^{\prime }-\theta }{2}-  \notag \\
-\frac{\eta (\theta ,t)}{R(t)}\left[ \eta (\theta ,t)v_{n}(\theta ,t)-\frac{1%
}{2\pi }\int\limits_{0}^{2\pi }\eta (\theta ^{\prime },t)v_{n}(\theta
^{\prime },t)d\theta ^{\prime }\right] +  \notag \\
+\frac{\eta ^{2}(\theta ,t)v_{n}(\theta ,t)}{R(t)}\frac{\partial }{\partial
\theta }\frac{1}{2\pi }\int\limits_{0}^{2\pi }d\theta ^{\prime }\ln \eta
(\theta ^{\prime },t)\cot \frac{\theta ^{\prime }-\theta }{2}-  \notag \\
-\frac{\eta (\theta ,t)}{R(t)}\frac{\partial }{\partial \theta }\frac{1}{%
2\pi }\int\limits_{0}^{2\pi }d\theta ^{\prime }v_{n}(\theta ^{\prime
},t)\eta (\theta ^{\prime },t)\cot \frac{\theta ^{\prime }-\theta }{2}
\label{4.1.30}
\end{gather}
considered in \cite{glb,gl1,gl2}. Here the radius $R\;$and$\,\ $normal
velocity $v_{n}\;$satisfy to the next equations: 
\begin{equation}
\frac{dR}{dt}=\frac{1}{2\pi }\int\limits_{0}^{2\pi }d\theta ^{\prime }\eta
(\theta ^{\prime },t)v_{n}(\theta ^{\prime },t).  \label{4.1.31}
\end{equation}

\begin{gather}
v_{n}(\theta ,\eta ,t)=\beta (\theta )\frac{\eta (\theta ,t)}{R(t)}\left\{
\varphi _{0}-\frac{\psi _{s}}{R(t)}\frac{\partial }{\partial \theta }\frac{1%
}{2\pi }\int\limits_{0}^{2\pi }d\theta ^{\prime }\cot \right. \times  \notag
\\
\times \frac{\theta ^{\prime }-\theta }{2}\eta (\theta ^{\prime
},t)d_{0}(\theta ^{\prime },\eta )\frac{\partial }{\partial \theta ^{\prime }%
}\left[ \frac{1}{2\pi }\int\limits_{0}^{2\pi }d\theta ^{\prime \prime }\cot 
\frac{\theta ^{\prime \prime }-\theta ^{\prime }}{2}\ln \eta (\theta
^{\prime \prime },t)\right] +  \notag \\
+\left. \frac{\varphi _{s}}{R(t)}\frac{\partial }{\partial \theta }\frac{1}{%
2\pi }\int\limits_{0}^{2\pi }d\theta ^{\prime }\cot \frac{\theta ^{\prime
}-\theta }{2}\eta (\theta ^{\prime },t)d_{0}(\theta ^{\prime },\eta
)\right\} .  \label{v}
\end{gather}

In local approximation last relationship has the form \cite{glb,gl1,gl2}: 
\begin{equation}
v_{n}(\theta ,\eta ,t)=\frac{\eta (\theta ,t)}{R(t)}\left( 1-\frac{1}{R(t)}%
\frac{\partial ^{2}\eta }{\partial \theta ^{2}}\right) .  \label{4.1.38}
\end{equation}

The equations (\ref{4.1.30}),(\ref{4.1.31})$,$(\ref{4.1.38}) are
one-dimensional equations determined on the interval $[0,2\pi ]$.

Now let us consider the linear stage of stability of a cylindrical boundary.
In this case it is natural to put $d_{0}\left( \theta ,\eta \right)
=d_{0}=const,\beta \left( \theta ,\eta \right) =\beta _{0}=const$ .
Linearizing equations (\ref{4.1.30})$,$ (\ref{4.1.31})$,$ (\ref{4.1.38})
with respect to perturbations $\delta \eta =\delta \eta _{0}\exp
(\int\limits_{0}^{t}\gamma \left( t^{\prime }\right) dt^{\prime }+ik\theta )$
at $k=\pm 1,\pm 2,...$ and taking into account the relation

\begin{equation*}
\frac{1}{2\pi }\int\limits_{0}^{2\pi }d\theta ^{\prime }\cot \frac{\theta
^{\prime }-\theta }{2}e^{ik\theta ^{\prime }}=i\ e^{ik\theta }\text{sign}k
\end{equation*}
we obtain the following characteristic equation 
\begin{equation}
\gamma =\frac{\beta }{R}\left[ k-2+\frac{d_{0}}{R}k^{2}(1-k)\right] .
\label{4.2.2}
\end{equation}

It can be seen from equation (\ref{4.2.2}) that the instability is realized
for rather short-wave fluctuations with $k>2$ \cite{MS63}. At $d_{0}\neq 0$
the fluctuations with wave number

\begin{equation}
k_{0}=\frac{1}{3}\left( \sqrt{1+3\left( \frac{d_{0}\varphi _{s}}{R\varphi
_{0}}\right) ^{-1}}-1\right)  \label{4.2.3}
\end{equation}
have the maximum increment of increase. The instability as $d_{0}\rightarrow
0$ is observed within a wide spectrum of wave vectors, which causes
complicated fractal structure of dendritic formations. The surface energy
damps the instability subject to chosen wave vectors and gives rise to the
minimum radius of curvature of a dendrite.

\section{Numerical simulation of the interface dynamics}

In the present section numerical simulation of the surface dynamics is
given. We consider the properties of the interface dynamics from the
viewpoint of singular integral equations.

\subsection{Properties of the interface dynamics}

Equations (\ref{4.1.30}) and (\ref{4.1.31}) are not directly related to the
physical mechanism determining motion of the interface in real space. They
determine the variation law governing the mapping of a circle into the
contour $\Gamma $ under the given normal velocity $v_{n}.$ In framework of
this approach we consider properties of equations (\ref{4.1.30}), (\ref
{4.1.31}).

We will investigate the periodical structure formation, supposing that the
structure will have period $2\pi /k,$ where $k$\ is an integer. This make it
possible to consider the solution only on the interval $\left[ 0,2\pi /k%
\right] $. In fact, we shall consider some integrable periodic function

\begin{equation}
\widetilde{f}(\theta +n2\pi /k)=\widetilde{f}(\theta )  \label{3.5.1}
\end{equation}
In this case the any singular integrals $J$ in the equation (\ref{4.1.30})
will look like:

\begin{eqnarray}
\frac{1}{2\pi }\int\limits_{0}^{2\pi }d\theta ^{\prime }\widetilde{f}(\theta
^{\prime })\cot \frac{\theta ^{\prime }-\theta }{2} &=&\sum_{n=0}^{k-1}\int%
\limits_{0}^{2\pi /k}d\theta ^{\prime }\widetilde{f}(\theta ^{\prime
})\left( \cot \frac{\theta ^{\prime }+n2\pi /k-\theta }{2}\right) = \\
&=&\int\limits_{0}^{2\pi /k}d\theta ^{\prime }\widetilde{f}(\theta ^{\prime
})\Re (\theta ^{\prime }-\theta )du^{\prime }
\end{eqnarray}
where

\begin{equation*}
\Re (u^{\prime }-u)=\sum_{n=0}^{k-1}\cot \frac{\theta ^{\prime }+n2\pi
/k-\theta }{2}.
\end{equation*}
So, we can use for computer simulation the interval [0,$2\pi /k$].

\subsection{Numerical solution of the interface dynamics}

The derived systems of nonlinear integro-differential equations were
investigated numerically. The numeric simulation was based on approximation
of unknown functions by trigonometrical polynomials. Expansions of this type
make it possible to use well-known quadrature formulae to evaluate singular
integrals. As a result, the system of equations (\ref{4.1.30}), (\ref{4.1.31}%
) and (\ref{4.1.38}) reduces to a system of ordinary non-linear differential
equations with filled Jacobian.

During computing experiments the approach based on approximations of
searched functions by trigonometrical polynomials \cite{6,pns}

\begin{equation}
\widetilde{f}_{n}(u)=\frac{1}{2\pi }\sum_{j=0}^{2n-1}\widetilde{f}%
(u_{j})\sin [n(u-u_{j})]\cot \frac{u-u_{j}}{2}  \label{3.5.6}
\end{equation}
with the given nodes set

\begin{equation}
u_{j}=\frac{\pi j}{n},\qquad j=0,1,...,2n-1.  \label{3.5.7}
\end{equation}
was used. The representation of the function $\widetilde{f}\left( u\right) $
in form (\ref{3.5.6}) makes it possible to apply the obtained analytical
expressions practically for all derivatives of the function $\widetilde{f}$
during calculations.

Besides, expansion of the function $\widetilde{f}\left( u\right) $ into a
trigonometrical polynomial makes it possible to use quadrature formulae \cite
{6,pns} for calculation of singular integrals with Hilbert kernels, that is

\begin{equation}
J(u_{m}^{0})=\sum_{j=0}^{2n-1}\frac{1}{2n}\widetilde{f}(u_{j})\cot \frac{%
u_{j}-u_{m}^{0}}{2}  \label{3.5.8}
\end{equation}
being defined on the system of nodes

\begin{equation}
u_{m}^{0}=\frac{2\pi m+1}{2\pi },\qquad m=0,1,...,2n-1.  \label{3.5.9}
\end{equation}

Taking now into account that all integrand functions on the right hand side
of equations (\ref{4.1.30}), (\ref{v}) are of the same kind it is possible
to present them in the form of (\ref{3.5.8}) making the following assignments

\begin{equation}
\eta _{m}(t,u)=\eta (t,u_{m}^{0}),\qquad m=0,1,...,2n-1.  \label{3.5.10}
\end{equation}
As a result from the equations of interface dynamics we obtain the system of
ordinary differential equations:

\begin{equation}
\frac{d\eta _{m}}{dt}=\widehat{F}_{m}\{\left\{ \eta
_{m},\sum_{j=0}^{2n-1}f_{j}\cot \frac{u_{j}-u_{m}^{0}}{2}\right\} ,
\label{3.5.11}
\end{equation}
where $\widehat{F_{m}}$ is a nonlinear operator being determined by the
right hand side of equations (\ref{4.1.30}), (\ref{4.1.31}) (\ref{v}), $\
f_{j}$ can take values $v_{n}(u_{j})\eta (u_{j}),\ln \eta (u_{j}),\eta
(u_{j})$ and etc.

Numerical integration of specified systems was carried by means of discrete
calculation of the Jacobian when realizing implicit methods of Girr and
Adams \cite{74}. Choice of the value of discretization and the local error
of integration was made so that the necessary accuracy of the searched
solution and the stability of calculations were ensured. Making use of
variable $\eta \left( u_{m}^{0}\right) $ changing in time, subject to the
transformations 
\begin{eqnarray}
dy &=&\frac{R(t)d\theta }{\eta (\theta ,t)}\sin \left( \theta +\psi (\theta
)+\frac{\pi }{2}\right) ,  \notag \\
dx &=&\frac{R(t)d\theta }{\eta (\theta ,t)}\cos \left( \theta +\psi (\theta
)+\frac{\pi }{2}\right)  \label{4.2.4}
\end{eqnarray}
for the system of equations (\ref{4.1.30}), (\ref{4.1.31}) and (\ref{4.1.38}%
) it is possible to reproduce geometrically defined interfaces in parametric
form.

In this text we evidently do not claim the complete investigation of pattern
formation in free boundary problems. Our goal is to show that developed
approach can be applied to investigation of interface evolution. Results of
computer simulation presented below give rise only to some sketch of the
structures realizing in the systems under consideration.

During numerical simulation evolution of the interface was investigated. The
product term $\beta $ affect only on time scale of the process. Therefore,
in the analysis under regard $\beta =1$ is taken. It was found out that the
surface energy influences the behavior of the field $\eta \left( \theta
,t\right) $ and interfaces too much. At small values of the surface energy
rather small-sized structures can be formed. As $d_{0}$ increases the
structures take a more smooth form. \FRAME{ftbpFU}{3.614in}{1.785in}{0pt}{%
\Qcb{Non-uniform distribution of the $\protect\eta (\protect\theta ,t),%
\protect\psi (\protect\theta ,t)$ and the corresponding interfaces for the
following initial perturbation: $\protect\eta (\protect\theta
,t_{0}=0)=1.0-0.01\cos (4\protect\theta );d_{0}=0.05;$ interfaces\ on the
figure correspond to the time steps $(t_{step}=0.5\ast 10^{4})$ $18$ times
are plotted, spaced between $0<t\leqslant 0.8\ast 10^{4}.$}}{\Qlb{rus2}}{%
rus2.gif}{\special{language "Scientific Word";type
"GRAPHIC";maintain-aspect-ratio TRUE;display "USEDEF";valid_file "F";width
3.614in;height 1.785in;depth 0pt;original-width 21.635in;original-height
10.6355in;cropleft "0";croptop "1";cropright "1";cropbottom "0";filename
'rus2.gif';file-properties "XNPEU";}}

Figure \ref{rus2} illustrates the characteristic form of non-uniform
distributions of the field $\eta $ (plot $a$) and the corresponding
interfaces (plot $b$). \FRAME{ftbpFU}{2.06in}{2.911in}{0pt}{\Qcb{Comparison
of interface profiles for different values of surface tention $d_{0}$ : $%
d_{0}=0.15-$a); $d_{0}=0.12-$b); $d_{0}=0.09-$c$)$; $d_{0}=0.06-$d$)$; $%
d_{0}=0.04-$e); $d_{0}=0.03-$f). Initial perturbations for all the figures
are the same: $\protect\eta (\protect\theta ,t_{0}=0)=1.0-0.01\cos (4\protect%
\theta );\;(t_{step}=0.5\ast 10^{4})$ 18 times are plotted, spaced between $%
0<t\leqslant 0.8\ast 10^{4}.$}}{\Qlb{rus1}}{rus1.gif}{\special{language
"Scientific Word";type "GRAPHIC";maintain-aspect-ratio TRUE;display
"USEDEF";valid_file "F";width 2.06in;height 2.911in;depth 0pt;original-width
13.1253in;original-height 18.6246in;cropleft "0";croptop "1";cropright
"1";cropbottom "0";filename 'rus1.gif';file-properties "XNPEU";}}

It is possible to observe that relatively smooth variations of the interface
correspond to rather great variations in the field $\eta $ as well as in the
angle $\psi .$ Thus, relatively small errors in determining $\eta $ do not
cause great changes in the interface geometry. In addition, as is seen from
Figures \ref{rus2} relatively even long sections of the interface shrink
when mapped onto the interval $\left[ 0,2\pi \right] $, whereas sharp
changes in the contour $\Gamma $, conversely, are stretched out. In this
sense such mappings are\ ``adaptive'' with respect to the information about
the detailed geometric structure of the interface.

\FRAME{ftbpFU}{1.9078in}{2.911in}{0pt}{\Qcb{Comparison of interface profiles
for different values of surface tention $d_{0}$: $d_{0}=0.15-a)$; $%
d_{0}=0.12 $-b); $d_{0}=0.09-$c); $d_{0}=0.06-$d); $d_{0}=0.04-$e); $%
d_{0}=0.03-$f). Initial perturbation for all the figures is the same: $%
\protect\eta (\protect\theta ,t_{0}=0)=1.0-0.01\cos (6\protect\theta
);(t_{step}=0.5\ast 10^{4})$ $18\;$times are plotted, spaced between $%
0<t\leqslant 0.8\ast 10^{4}.$}}{\Qlb{iz6}}{izotrop_group_6.gif}{\special%
{language "Scientific Word";type "GRAPHIC";maintain-aspect-ratio
TRUE;display "USEDEF";valid_file "F";width 1.9078in;height 2.911in;depth
0pt;original-width 13.677in;original-height 20.9691in;cropleft "0";croptop
"1";cropright "1";cropbottom "0";filename
'izotrop_group_6.gif';file-properties "XNPEU";}}

The characteristic geometry of the structures obtained as a result of
computer simulation of a small perturbation is presented on Figures \ref
{rus1}, and \ref{iz6}. The smaller capillary length is the more diverse
structure arises. Figures \ref{rus1}, and \ref{iz6} are different from each
other due to periodicity of the initial perturbation.

If the system is influenced by an uncorrelated noise with the normal law of
distribution unregular structures arise. During simulation it was taken that
on each step of integration the velocity at point $i$ of the interval $\left[
0,2\pi \right] $ equal to $v_{n}\left( \theta _{i}\right) +\left( \eta
/R\right) ^{1/2}\delta v_{n}\left( \theta _{i}\right) $, within $\delta
v_{n}\left( \theta _{i}\right) $ being chosen in the correspondence with the
normal law of distribution in limits $\left( -10^{-5},10^{-5}\right) $ (Fig. 
\ref{stah}).

\FRAME{ftbpFU}{1.0438in}{1.0283in}{0pt}{\Qcb{Interface dynamics stimulated
by external noise. }}{\Qlb{stah}}{stochast.gif}{\special{language
"Scientific Word";type "GRAPHIC";maintain-aspect-ratio TRUE;display
"USEDEF";valid_file "F";width 1.0438in;height 1.0283in;depth
0pt;original-width 2.9378in;original-height 2.8963in;cropleft "0";croptop
"1";cropright "1";cropbottom "0";filename 'Stochast.gif';file-properties
"XNPEU";}}

\FRAME{ftbpFU}{3.7282in}{1.8775in}{0pt}{\Qcb{Non-uniform distribution of the 
$\protect\eta (\protect\theta ,t),\protect\psi (\protect\theta ,t)$ and the
dendrite silhouette for surface tension anizotropy: $\protect\eta (\protect%
\theta ,t_{0}=0).=1-0.01\cos (4\protect\theta );d_{0}=0.3;$ $\protect\gamma %
_{4}=0.05$ interfaces\ on the figure correspond to the time steps$%
(t_{step}=0.25\ast 10^{4})$ 15 times are plotted, spaced between $%
0<t\leqslant 0.325\ast 10^{5}.$}}{\Qlb{rus4}}{rus4an.gif}{\special{language
"Scientific Word";type "GRAPHIC";maintain-aspect-ratio TRUE;display
"USEDEF";valid_file "F";width 3.7282in;height 1.8775in;depth
0pt;original-width 21.3643in;original-height 10.7081in;cropleft "0";croptop
"1";cropright "1";cropbottom "0";filename 'rus4an.gif';file-properties
"XNPEU";}}

For simulation of anisotropic growth of crystals the following expression
(similar as in \cite{115}) was used:

\begin{gather}
d_{0}(\theta )=(1-15\gamma _{4}\cos 4\gamma )d_{0}=  \notag \\
=\left[ 1-15\gamma _{4}\cos 4\left( \theta -\frac{1}{2\pi }%
\int\limits_{0}^{2\pi }d\theta ^{\prime }\ln \eta (\theta ^{\prime })\cot 
\frac{\theta ^{\prime }-\theta }{2}\right) \right] d_{0}.  \label{4.2.5}
\end{gather}

The interfaces corresponding to the capillary length (\ref{4.2.5}), subject
to the anisotropy of surface energy have the form of truly looking cross
(Fig. \ref{rus4}, and \ref{rus3}), despite that at the initial moment the
boundary surface presents a circle (up to $\delta v_{n}|_{t=0}\cong 10^{-5})$%
. According to the value of $\gamma _{4}$ such an anisotropy is exhibited at
small or large values of $R$ and $t.$ The realized distributions little
depend on initial perturbations and reduce frequently to rather complicated
and exotic surfaces. A more developed structure in the case of anisotropy is
presented by Fig. \ref{rus7}. An example of structures arising from
nonhomogeneous initial conditions is illustrated by Fig. \ref{ad1}, $a)$. In
the same time Fig. \ref{ad1}, $b)$ illustrates an example of structures
developing from a flat boundary.

\FRAME{ftbpFU}{2.9144in}{2.6541in}{0pt}{\Qcb{Comparison of interfacee
profiles for different values of surface tension anizotropy: $d_{0}=0.3;$ a)$%
\protect\gamma _{4}$=$0.00$; b)$\protect\gamma _{4}$=$0.01$ c); $\protect%
\gamma _{4}$=$0.02$; d)$\protect\gamma _{4}$=$0.04.$ Initial perturbation
for all the figures is the same: $\protect\eta (\protect\theta
,t_{0}=0)=1.-0.01\cos (4\protect\theta )$; $(t_{step}=0.25\ast 10^{4})$ 15
times are plotted, spaced between $0<t\leqslant 0.325\ast 10^{5}.$}}{\Qlb{%
rus3}}{rus3.gif}{\special{language "Scientific Word";type
"GRAPHIC";maintain-aspect-ratio TRUE;display "USEDEF";valid_file "F";width
2.9144in;height 2.6541in;depth 0pt;original-width 15.7915in;original-height
14.3749in;cropleft "0";croptop "1";cropright "1";cropbottom "0";filename
'rus3.gif';file-properties "XNPEU";}}

It should be noted that at $d_{0}=0$ the results concerning smooth boundary
under stochastic excitation of short-wave perturbations, caused for example
by errors of calculations do not lead the developed structures due to
instabilities of computation.

\FRAME{ftbpFU}{2.1223in}{2.1127in}{0pt}{\Qcb{Dendrite silhoette for surface
tension anizotropy: $\protect\gamma =0.03,\;d_{0}=0.03$, $\protect\eta (%
\protect\theta ,t_{0}=0)=1-0.01\exp \left( \frac{\protect\theta -\protect\pi
/4}{0.1}\right) ^{2}$, $(t_{step}=0.2\ast 10^{4})$ 21 times are plotted,
spaced between $0<t\leqslant 0.36\ast 10^{4}.$}}{\Qlb{rus7}}{rus7.gif}{%
\special{language "Scientific Word";type "GRAPHIC";maintain-aspect-ratio
TRUE;display "USEDEF";valid_file "F";width 2.1223in;height 2.1127in;depth
0pt;original-width 4.625in;original-height 4.6043in;cropleft "0";croptop
"1";cropright "1";cropbottom "0";filename 'rus7.gif';file-properties
"XNPEU";}}

\FRAME{ftbpFU}{2.9464in}{1.8697in}{0pt}{\Qcb{Dendrite silhoette for surface
tension anizotropy: $\protect\gamma =0.00,\;d_{0}=0.03$, $\protect\eta (%
\protect\theta ,t_{0}=0)=1-0.01\exp \left( \frac{\protect\theta -\protect\pi
/4}{0.1}\right) ^{2}$, $(t_{step}=0.1\ast 10^{4})$ 13 times are plotted,
spaced between $0<t\leqslant 0.12\ast 10^{4}.-a).$ The evolution of the flat
boundary with small initial perturbation $\protect\eta \left( t=0\right)
=1-0.01\cos u$ $-b)$.}}{\Qlb{ad1}}{ad1.gif}{\special{language "Scientific
Word";type "GRAPHIC";maintain-aspect-ratio TRUE;display "USEDEF";valid_file
"F";width 2.9464in;height 1.8697in;depth 0pt;original-width
4.1355in;original-height 2.6143in;cropleft "0";croptop "1";cropright
"1";cropbottom "0";filename 'ad1.gif';file-properties "XNPEU";}}

\end{document}